\documentstyle[12pt]{article}

\newcommand{\be}{\begin{equation}}
\newcommand{\ee}{\end{equation}}
\def\bea{\begin{eqnarray}}
\def\eea{\end{eqnarray}}
\def\nn{\nonumber \\}
\def\qqg{$q {\bar q}g$}
\def\qq{$q {\bar q}$}
\def\pp{$p_{\perp}$}

\input psfig

\newcommand{\gnufig}[3]{
\begin{figure}[t]
\begin{center}
{#3}
\end{center}
\caption{#2}
\label{fig:#1}
\end{figure}}

\begin{document}
\date{}
\title{
{\large\rm DESY 97-099}\hfill{\large\tt ISSN 0418-9833}\\
{\large\rm DAMTP-97-59}\hfill\vspace*{0cm}\\
{\large\rm June 1997}
\hfill\vspace*{2.5cm}\\
High-$p_\perp$ Jets in Diffractive Electroproduction}
\author{W. Buchm\"uller, M. F. McDermott\\
{\normalsize\it Deutsches Elektronen-Synchrotron DESY, 22603 Hamburg, 
Germany}
\\[.2cm]
and\\[.2cm]
A. Hebecker\\
{\normalsize\it D.A.M.T.P., Cambridge University, Cambridge CB3 9EW, 
England}
\vspace*{2cm}\\                     
}

\maketitle  
\begin{abstract}
\noindent
The diffractive production of high-$p_{\perp}$ jets in deep-inelastic 
scattering is studied in the semiclassical approach. The $p_{\perp}$-spectra 
of \qq\ and \qqg\ diffractive final states are found to be qualitatively 
different. For \qq\ final states, which are produced by `hard' 
colour-singlet exchange, the $p_{\perp}$-spectrum is much softer than for 
\qqg\ final states, where the colour neutralization is `soft'. Furthermore, 
the two different final states can be clearly distinguished by their 
diffractive mass distributions.
 
\end{abstract} 
\thispagestyle{empty}
\newpage                                             
The observation of high-$p_{\perp}$ jets in diffractive electron-proton
scattering at small $x$ \cite{hera} provides direct evidence for an
intriguing interplay of `soft' and `hard' interactions in QCD. To disentangle
these two aspects of diffractive deep-inelastic scattering detailed studies of different
final states will be important.

One might expect the hard scale, provided by the transverse momentum of the
jets, to ensure the applicability of perturbation theory. Indeed, the 
production of final states containing only two high-$p_\perp$ jets can be 
described by perturbative two-gluon exchange \cite{mue}. This process has 
been studied in detail by several groups and higher-order corrections have 
already partially been considered \cite{nz}.  

In this paper we study the diffractive production of high-$p_{\perp}$ jets 
in the semiclassical approach \cite{bh,bhm}. In the proton rest frame, one 
calculates the high energy scattering of  partonic fluctuations of the 
virtual photon by the colour field of the proton. We shall consider the two 
simplest configurations, \qq\ and  \qqg. In both cases  diffractive 
processes are obtained by projecting onto the colour singlet configuration of 
the final state partons.

The production of high-$p_\perp$ jets is dominated by \qqg\ final states 
where the gluon has low transverse momentum and carries a small energy 
fraction of the photon \cite{bhm}. Kinematically, the presence of such a 
`wee' parton in the photon implies that the fluctuation develops a large 
transverse size by the time it reaches the proton. The corresponding hard 
process is boson-gluon fusion \cite{h}. We also calculate the 
$p_{\perp}$-spectrum for the \qq\ final state. With a proper identification 
of the inclusive gluon density one obtains the same result as two-gluon 
exchange in leading order. The two different final states can be directly 
distinguished by measuring the invariant mass of the two high-$p_{\perp}$ 
jets.

Our analysis is similar to the case of diffractive charm considered in 
\cite{bhm2}. Note however that in the present paper the production of both 
\qq\ and \qqg\ final states is described exclusively within the 
semiclassical approach.\\

\noindent
{\large\bf Hard and soft colour neutralization}\\

In the semiclassical approach inclusive and diffractive cross sections can 
be expressed at leading order in terms of a single non-perturbative 
quantity, $\mbox{tr}\,W^{\cal F}_{x_{\perp}}(y_{\perp})$, where 
\be
W^{\cal F}_{x_\perp}(y_\perp)=U^\dagger(x_\perp+y_\perp)U(x_\perp)-1\,
\label{wa}
\ee
is built from the non-Abelian eikonal factors $U$ and $U^\dagger$ of
the quark 
and antiquark whose light-like paths penetrate the colour field of the 
proton at transverse positions $x_\perp$ and $x_\perp+y_\perp$, respectively 
(cf.~Fig.~1). The superscript ${\cal F}$  is 
used because the quarks are in the fundamental representation of the gauge 
group. Since the colour field outside the proton vanishes 
$W^{\cal F}_{x_{\perp}}(y_{\perp})$ is essentially a closed Wilson loop 
through a section of the proton which measures an average of the proton 
colour field.

\begin{figure}[ht]
\begin{center}
\vspace*{-.5cm}
\parbox[b]{11cm}{\psfig{width=10cm,file=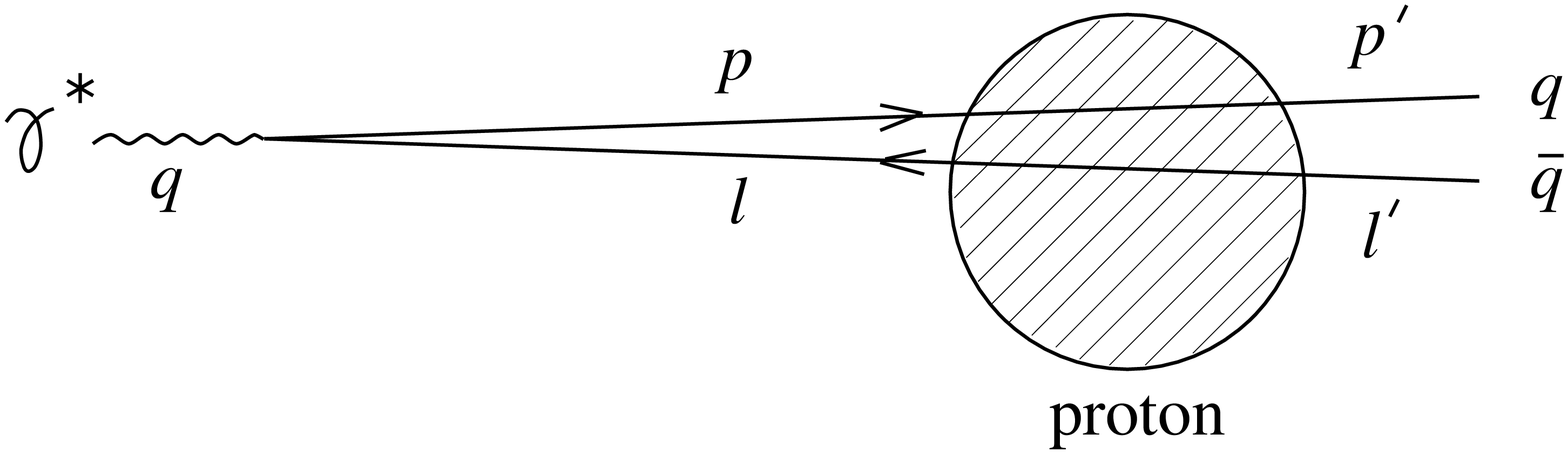}}\\
\end{center}
\refstepcounter{figure}
\label{qq}
Figure \ref{qq}: Exclusive two-jet production in the semiclassical approach.
\end{figure}

In an expansion in the transverse distance between quark and antiquark
one has
\be\label{wsh}
\int_{x_{\perp}}\mbox{tr}W^{\cal F}_{x_{\perp}}(y_{\perp})
= - {1\over 4} y_{\perp}^2 C_1 + {\cal O}(y_{\perp}^4)\, .\label{intw}
\ee
Using the unitarity of the matrix $U$ one can easily derive the constant 
\be
C_1 = \int_{x_{\perp}} \mbox{tr}\left(\partial_{y_{\perp}}
W^{\cal F}_{x_{\perp}}(0)\partial_{y_{\perp}}
W_{x_{\perp}}^{{\cal F} \dagger}(0)\right)\, .
\ee
In the semiclassical approach this quantity determines the variation of the 
inclusive structure function $F_2(x,Q^2)$ with $Q^2$ \cite{bhm}. A 
comparison with  boson-gluon fusion in the parton model yields 
\be\label{glue}
C_1 = 2 \pi^2 \alpha_s x G(x)\, 
\ee
for the inclusive gluon density. Since $C_1$ is constant $G(x) \sim 1/x$
which corresponds to a classical bremsstrahl spectrum of gluons.

The integral over $\mbox{tr}W^{\cal F}_{x_{\perp}}(y_{\perp})$ in 
Eq.~(\ref{intw}) describes the cross section, $\sigma(y_\perp)$, for the 
scattering of a colour dipole of transverse size $y_{\perp}$ off the proton, 
within the semiclassical approach. A corresponding relationship between 
$\sigma(y_\perp)$ and the inclusive gluon density is given in \cite{fms}. 

We now consider the production of a diffractive \qq\ final state. Using the 
results of \cite{bhm} (cf. Appendix B) for the transversely polarized photon 
one easily finds
\bea\label{hse}
\!\!\!\!\!
\left.{d\sigma_T\over dt}\right|_{t=0} &\!\!\!\!\!\!=&\!\!\! {\sum_q e_q^2 
 \alpha_{em} \over 6(2\pi)^6} \!\!\int\!\! d\alpha dl'^2_{\perp} (\alpha^2 + 
 (1\!-\!\alpha)^2) \left| \int_{x_{\perp},y_{\perp},l_{\perp}} e^{iy_{\perp}
 (l'_{\perp}-l_{\perp})} \mbox{tr}W^{\cal F}_{x_{\perp}}(y_{\perp})
 {l_{\perp}\over a^2 + l_{\perp}^2} \right|^2\!\!\! , 
\\ \nonumber\\
 && \hspace{2cm} a^2=\alpha (1-\alpha)Q^2\, .
\eea
Here $t=(q-p'-l')^2$ is the momentum transfer to the proton and $\alpha=l'_0
/q_0$. The cross section for large transverse momenta is dominated by the 
short distance behaviour of $\mbox{tr}W^{\cal F}_{x_{\perp}}(y_{\perp})$. 
Inserting Eq.~(\ref{wsh}), substituting $y_{\perp}^2 \rightarrow - (\partial/
\partial l_{\perp})^2$ and using $l'_{\perp}\simeq -p'_{\perp}$, one obtains 
\be\label{xqq}
\left.{d\sigma_T \over dt d\alpha dp'^2_{\perp}}\right|_{t=0}=
{\sum_q e_q^2 \alpha_{em} C_1^2\over 384\pi^2} (\alpha^2 + (1-\alpha)^2)
\left|\left({\partial\over \partial p'_{\perp}}\right)^2
{p'_{\perp}\over a^2 + p'^2_{\perp}}\right|^2\, .
\ee
As the derivation illustrates, this cross section describes the interaction
of a small \qq\ pair with the proton. Hence, it is perturbative or hard.
According to Eq.~(\ref{glue}) the cross section Eq.~(\ref{xqq}) is 
proportional to the square of the gluon density. In order to obtain the 
integrated cross section one has to multiply Eq.~(\ref{xqq}) by the constant 
\be
C=\left(\int\frac{d\sigma}{dt}dt\right)\bigg/\left(\frac{d\sigma}{dt}
\bigg|_{t\approx0}\right)\sim \Lambda^2\, ,
\ee
where $\Lambda$ is a typical hadronic scale. The resulting cross section
integrated down to the transverse momentum 
$p'^2_{\perp,\mbox{\footnotesize cut}}$ yields a contribution to the 
diffractive structure function $F_2^D$ which is suppressed by 
$\Lambda^2/p'^2_{\perp,\mbox{\footnotesize cut}}$. 

As shown in \cite{bhm}, a `leading twist' cross section with jets of 
$p_{\perp}\sim Q$ requires at least three partons in the final state, one of 
which has low transverse momentum. The corresponding cross sections can be 
written as convolution of ordinary partonic cross sections with diffractive 
parton densities \cite{sop}. In the case of high-$p_{\perp}$ quark jets 
there is an additional wee gluon (cf.~Fig.~2). The 
partonic process is then boson-gluon fusion and the cross section 
\be\label{bgfu}
{d\sigma_T\over d\xi dp'^2_{\perp}} = \int_x^\xi dy 
{d\hat{\sigma}_T^{\gamma^* g\rightarrow q\bar{q}}(y,p_{\perp}') 
 \over dp'^2_{\perp}} {dg(y,\xi)\over d\xi}
\ee
involves a diffractive gluon density \cite{h}
\be
{dg(y,\xi)\over d\xi} = {1\over 8\xi^2} \left(b\over 1-b\right)\int {d^2 
k'_\perp (k'^2_\perp)^2 \over (2\pi)^4}\int_{x_\perp}\left|\int {d^2 
k_\perp\over (2\pi)^2} {\mbox{tr}[\tilde{W}^{\cal A}_{x_\perp}(k'_\perp-
k_\perp)]t^{ij}\over k_\perp'^2 b+k_\perp^2 (1-b)}\right|^2\, ,
\ee
\be
t^{ij}=\delta^{ij} + {2k^i_{\perp}k^j_{\perp}\over k'^2_{\perp}}
\left({1-b\over b}\right)\, .
\ee
Here $\xi=x (Q^2+M^2)/Q^2$ for a final state with diffractive mass $M$, the 
momentum fraction of the proton carried by the incoming gluon is denoted by 
$y$, and $b=y/\xi$. The function $\tilde{W}^{\cal A}_{x_\perp}$ is the 
Fourier transform of $W^{\cal A}_{x_\perp}$ which is defined as in 
Eq.~(\ref{wa}) but with the $U$-matrices in the adjoint
representation, hence the superscript ${\cal A}$.

\begin{figure}[ht]
\begin{center}
\vspace*{-.5cm}
\parbox[b]{11cm}{\psfig{width=10cm,file=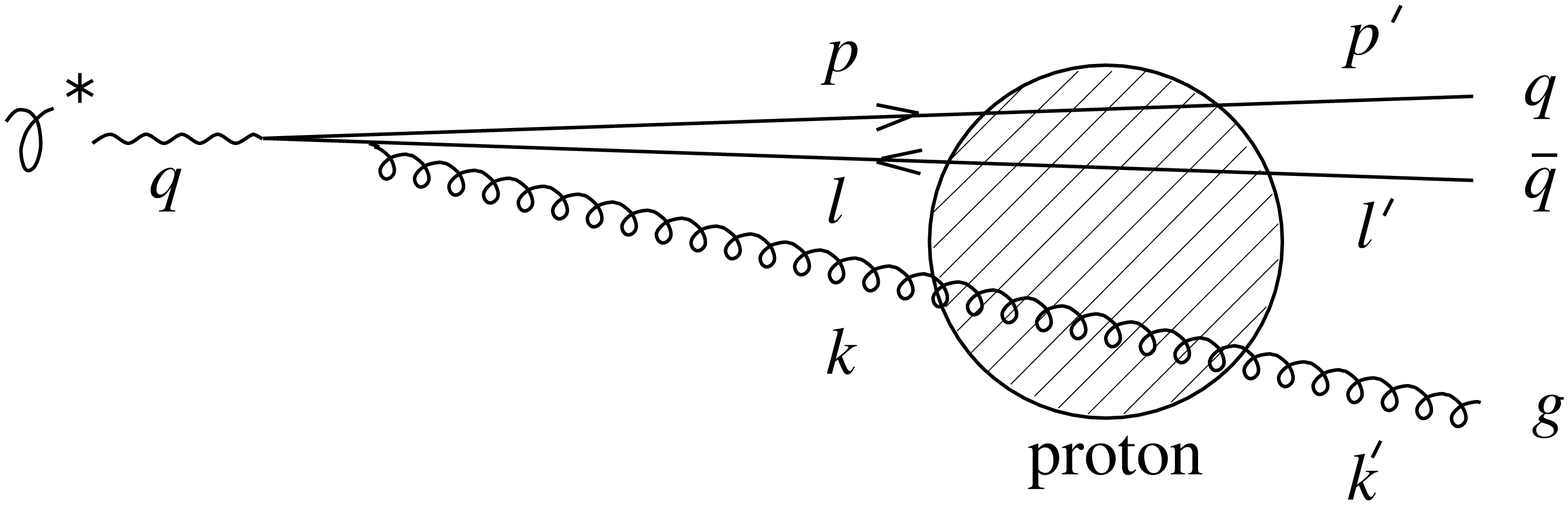}}\\
\end{center}
\refstepcounter{figure}
\label{qqg}
Figure \ref{qqg}: Two-jet production with an additional low transverse
momentum gluon.
\end{figure}

\begin{figure}[ht]
\begin{center}
\vspace*{-.5cm}
\parbox[b]{10.5cm}{\psfig{width=9.5cm,file=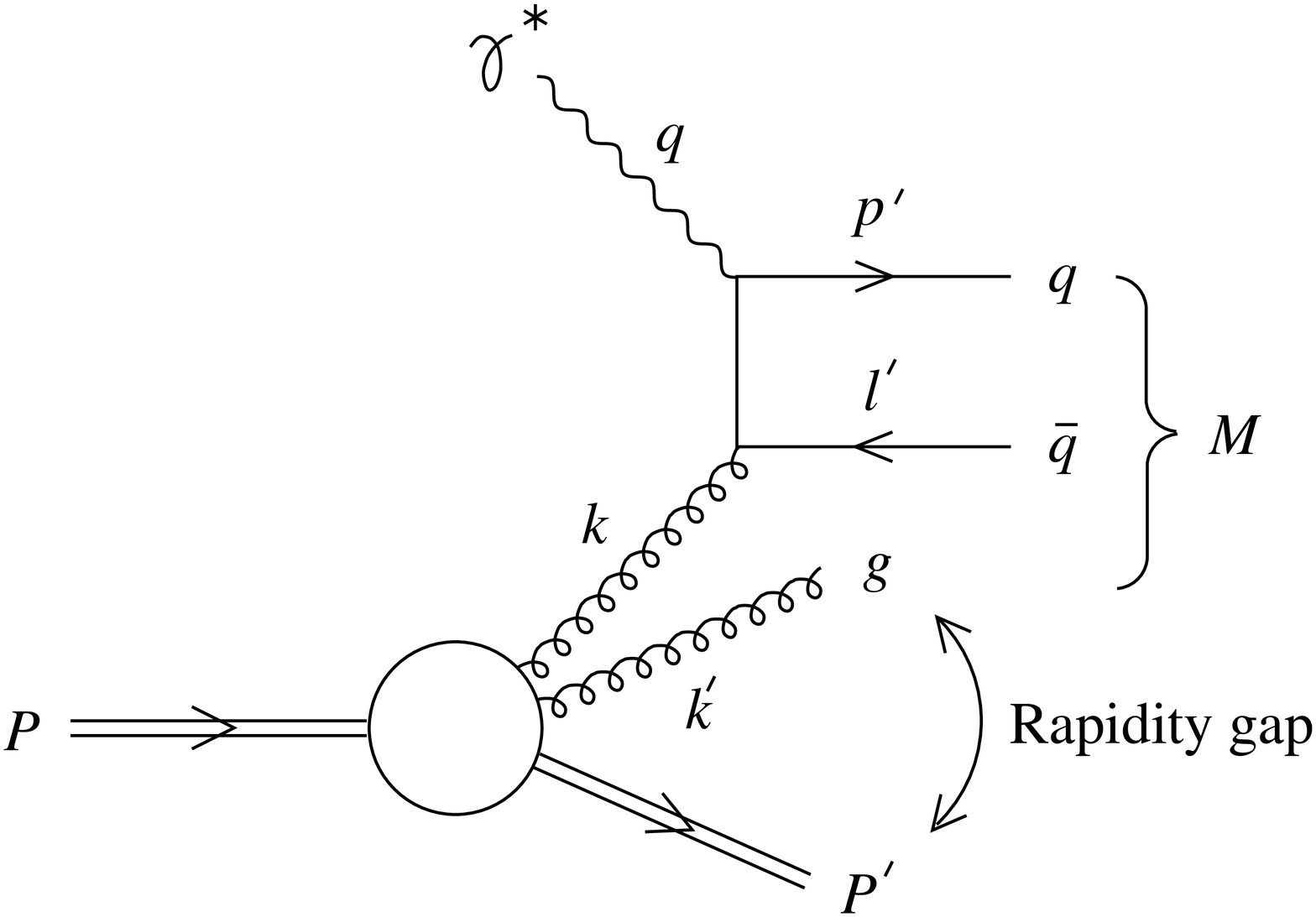}}\\
\end{center}
\refstepcounter{figure}
\label{bgf}
Figure \ref{bgf}: Interpretation of the process of Fig.~\ref{qqg}
in terms of boson-gluon fusion in a frame where the proton is fast, e.g.,
the Breit frame.
\end{figure}

Eq.~(\ref{bgfu}) is formulated in the partonic language which is
appropriate in a frame in which the proton is fast. 
The corresponding process is shown in Fig.~\ref{bgf}: a 
low transverse momentum colour-singlet pair of gluons is extracted from the 
proton, which is elastically scattered. One of the gluons, corresponding to 
the initial state off-shell gluon in Fig.~\ref{qqg}, produces a 
high-$p_{\perp}$ pair of quark jets via boson-gluon fusion. Hence, the hard 
process is the same as in the diffractive parton model proposed in 
\cite{bh0}. However, in contrast to this model, the semiclassical approach 
predicts an additional low transverse momentum gluon in the final state,
which reflects the non-perturbative mechanism of colour neutralization. 

In addition to boson-gluon fusion, the QCD Compton process can also produce
high-$p_{\perp}$ jets. In this case either the quark or the antiquark
is the wee parton. The corresponding cross section can be expressed in terms 
of a diffractive quark density \cite{h},
\bea\label{com}
{d\sigma_T\over d\xi dp'^2_{\perp}} &=& \int_x^\xi dy 
{d\hat{\sigma}_T(y,p'_{\perp})^{\gamma^*q\rightarrow gq} 
\over dp'^2_{\perp}}{dq(y,\xi)\over d\xi}\, ,\\
{dq(y,\xi)\over d\xi} &=& {2\over 3\xi^2} \int {d^2 l'_\perp (l'^2_\perp) 
\over (2\pi)^4}\int_{x_\perp}\left|\int {d^2 l_\perp\over (2\pi)^2} {l_\perp
\mbox{tr} [\tilde{W}^{\cal F}_{x_\perp}(l'_\perp-l_\perp)]\over l_\perp'^2 b+
l_\perp^2 (1-b)}\right|^2\, .
\eea

\noindent An analogous relation holds in the antiquark case.

The diffractive quark and gluon densities depend on $\mbox{tr}
\tilde{W}_{x_{\perp}}$ which is a function of the transverse momentum 
transfer to the wee parton. The integration is dominated by small 
values of this momentum transfer (of order $\Lambda^2$). Hence, the 
diffractive quark and gluon densities are non-perturbative quantities which 
describe a soft interaction of the virtual photon with the proton. 

We will not discuss the energy dependence of the above contributions 
to the diffractive jet cross section. For a given diffractive mass the 
energy dependence in the semiclassical approach is flat in both cases. 
However, this dependence will be increased by higher-loop corrections.\\

\noindent
{\large\bf Transverse momentum distribution of diffractive jets}\\ 
\nopagebreak

The cross sections of Eqs.~(\ref{bgfu}), (\ref{com}) for diffractive boson-gluon
fusion and diffractive Compton scattering, respectively, can be evaluated
along the lines described in \cite{bhm2}. Therefore, in the following, we just 
list the relevant results. In the leading-$\ln(1/x)$ approximation one 
obtains for the longitudinal and transverse boson-gluon fusion cross sections 
\begin{eqnarray}
\frac{d\sigma_L}{d\alpha dp_\perp'^2}&=&\frac{\Sigma_q e_q^2\alpha_{em}
\alpha_s}{2\pi^3}\, \frac{[\alpha(1-\alpha)]^2Q^2p_\perp'^2}{(a^2+p_\perp'^2
)^4}\, \ln(1/x) h_{\cal A}, \label{eq:slg}
\\ \nn
\frac{d\sigma_T}{d\alpha dp_\perp'^2}&=&\frac{\Sigma_q e_q^2\alpha_{em}
\alpha_s}{16\pi^3}\,\frac{(\alpha^2+(1\!-\!\alpha)^2)\,(p_\perp'^4+a^4)}
{(a^2+p_\perp'^2)^4}\,\ln(1/x) h_{\cal A} \,,\label{eq:stg} 
\\
h_{\cal A} &=&\int_{y_\perp}\int_{x_\perp}\frac{\left|\mbox{tr}
W^{\cal A}_{x_\perp}(y_\perp)\right|^2}{y_\perp^4}\,.\label{ha}
\end{eqnarray}
Similarly, one finds for the QCD-Compton cross sections
\begin{eqnarray}
\frac{d\sigma_L}{d\alpha dp_\perp'^2}&=&\frac{16 \Sigma_q e_q^2\alpha_{em}
\alpha_s}{27\pi^3} \frac{Q^2}{[\alpha(1-\alpha)] N^6}\, h_{\cal F}, 
\label{eq:slq}
\\ \nonumber\\
\frac{d\sigma_T}{d\alpha dp_\perp'^2}&=&\frac{4 \Sigma_q
  e_q^2\alpha_{em}\alpha_s}{27 \pi^3 N^6 p_\perp'^2} 
\left[ N^4 - 2Q^2(N^2+Q^2) +
  \frac{N^4+Q^4}{\alpha(1\!-\!\alpha)} \right]
\,h_{\cal F}\, , \label{eq:stq}\\
&&\qquad N^2=Q^2 + {p'^2_{\perp}\over \alpha(1-\alpha)}\, ,
\end{eqnarray}
where the constant $h_{\cal F}$ is defined analogously to Eq.~(\ref{ha}).

Comparing Eqs.~(\ref{eq:slg}), (\ref{eq:stg}) with Eqs.~(\ref{eq:slq}), 
(\ref{eq:stq}) we see that the configurations with a 
wee gluon are enhanced by $\ln(1/x)$ at small $x$ relative to those 
with a wee quark or antiquark. The simple arguments concerning 
colour outlined in \cite{bhm2} suggest an additional large suppression of 
the wee fermion contributions due to colour factors ($h_{\cal A} \approx 
16 h_{\cal F}$). As a result we claim that the configurations with a wee 
gluon dominate over those with a wee fermion in the small-$x$ region 
relevant to diffraction and we shall ignore the latter from now on.

We can also calculate the differential cross section for the leading order 
\qq\ fluctuation, as described in the previous section (cf. 
Eq.~(\ref{xqq})). The longitudinal and transverse cross sections are
\begin{eqnarray}
\frac{d\sigma_L}{d\alpha dp'^{2}_\perp}&=&\frac{2\Sigma_q e_q^2\alpha_{em}
\alpha_s^2\pi^2[\xi G(\xi)]^2C}{3}\, \frac{[\alpha(1-\alpha)]^2Q^2 (a^2-
p'^{2}_\perp)^2}{(a^2+p'^{2}_\perp)^6}\,,\label{slg} 
\\ \nonumber\\
\frac{d\sigma_T}{d\alpha dp'^{2}_\perp}&=&\frac{2 \Sigma_q e_q^2\alpha_{em}
\alpha_s^2\pi^2[\xi G(\xi)]^2C}{3} \,\frac{(\alpha^2+(1-\alpha)^2)
p'^{2}_\perp a^4}{(a^2+p'^{2}_\perp)^6}\, .\label{eq:stqqg}
\end{eqnarray}
\noindent Identical differential distributions have been found for two-gluon
exchange in leading order \cite{nz}. 
One can easily see that the region of $\alpha$ close to zero or one 
dominates and that high-\pp  configurations are unlikely. We now attempt to 
quantify this statement.

The quantitative differences between the \qq\ and \qqg\ configurations are 
particularly  pronounced in the integrated cross section with a lower cut on 
the  transverse  momentum of the quarks. Since the overall normalization of the
contributions is uncertain (it is inherently non-perturbative) we compare 
the shape in $p_{\perp}'^2$ of each configuration by considering the 
quantity
\be
\sigma(p'^2_{\perp,\mbox{\footnotesize cut}})= 
\int_{p'^2_{\perp,\mbox{\scriptsize cut}}}^{\infty} {dp_\perp'^2}
\int_{0}^{1} d\alpha
\frac{d\sigma}{dp_\perp'^2 d\alpha}
\ee
which is the fraction of events remaining above a certain minimum 
$p_\perp'^2$. The integrand here is obtained by adding the contributions 
from longitudinal and transverse photons, in each case. 
Fig.~\ref{fig:psq} shows the dependence of the corresponding event
fraction on the lower limit, $p'^2_{\perp,\mbox{\footnotesize cut}}$. 
Each curve is normalized to its value at 
$p'^2_{\perp,\mbox{\footnotesize cut}} = 5 $~GeV$^2$. One can see that the 
spectrum for the \qqg\ configuration is much harder than that for the \qq\ 
configuration. This is expected since in boson-gluon fusion $p_\perp$ is 
distributed logarithmically between the soft scale and $Q$ thus resulting 
in a significant high-$p_\perp$ tail above 
$p'_{\perp,\mbox{\footnotesize cut}}$.\\ 

\gnufig{psq}{The fraction of diffractive events with $p'^2_{\perp}$ above 
$p'^2_{\perp,\mbox{\footnotesize cut}}$ for $Q^2$ of 10 GeV$^2$ and 100
GeV$^2$ (lower and upper curve in each pair).}{
\setlength{\unitlength}{0.1bp}
\begin{picture}(4320,2592)(0,0)
\includegraphics{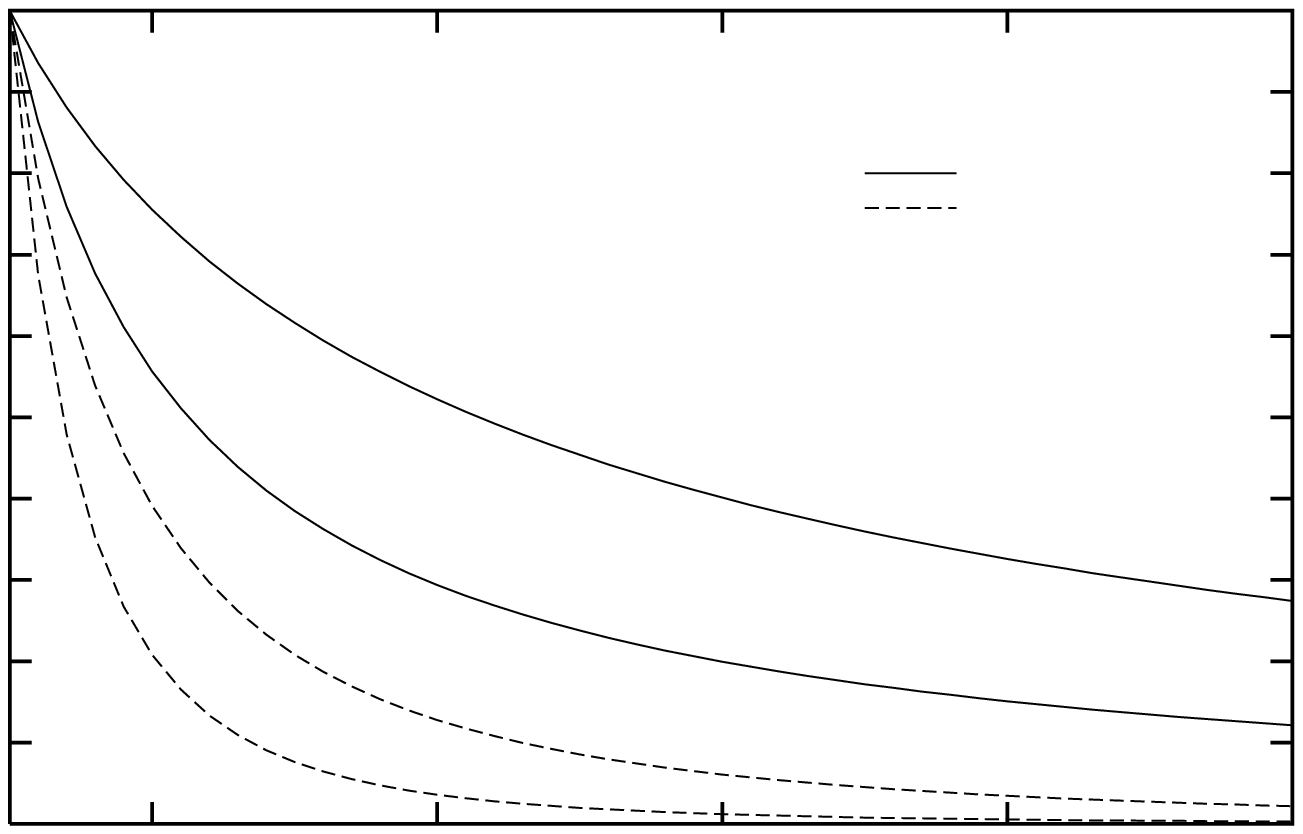}
\put(2876,1924){\makebox(0,0)[r]{ }}
\put(2876,2024){\makebox(0,0)[r]{ }}
\put(2679,1907){\makebox(0,0)[l]{$q \bar{q} ~~ $}}
\put(2679,2024){\makebox(0,0)[l]{$q \bar{q} g $}}
\put(2023,-118){\makebox(0,0)[l]{$p'^2_{\perp,\mbox{cut}} ~(\mbox{GeV}^2)$}}
\put(100,1321){%
\makebox(0,0)[b]{\shortstack{Fraction  of Events}}%
}
\put(4157,50){\makebox(0,0){50}}
\put(3336,50){\makebox(0,0){40}}
\put(2515,50){\makebox(0,0){30}}
\put(1694,50){\makebox(0,0){20}}
\put(873,50){\makebox(0,0){10}}
\put(413,2492){\makebox(0,0)[r]{1}}
\put(413,2258){\makebox(0,0)[r]{0.9}}
\put(413,2024){\makebox(0,0)[r]{0.8}}
\put(413,1789){\makebox(0,0)[r]{0.7}}
\put(413,1555){\makebox(0,0)[r]{0.6}}
\put(413,1321){\makebox(0,0)[r]{0.5}}
\put(413,1087){\makebox(0,0)[r]{0.4}}
\put(413,853){\makebox(0,0)[r]{0.3}}
\put(413,618){\makebox(0,0)[r]{0.2}}
\put(413,384){\makebox(0,0)[r]{0.1}}
\put(413,150){\makebox(0,0)[r]{0}}
\end{picture}
}

\noindent
{\large\bf Mass distribution of the diffractive jet system}\\ \nopagebreak

Let $M_j$ be the invariant mass of the two-jet system in diffractive events 
containing two high-$p_\perp$ jets in the diffractive final state. The 
measurement of this observable provides, in principle, a clean distinction 
between \qq\ final states, where $M_j^2=M^2$, and \qqg\ final states, where 
$M_j^2<M^2$. In practice, however, this requires the contribution of the 
wee gluon to the diffractive mass, which is responsible for the difference 
between $M^2$ and $M_j^2$, to be sufficiently large. To quantify the 
expectation within the semiclassical approach we consider the transverse 
photon contribution to the differential diffractive cross section 
$d\sigma/dM^2dM_j^2$. 

In the case of a \qq\ final state this cross section can be obtained 
directly from Eq.~(\ref{eq:stqqg}),
\be
\frac{d\sigma_T}{dM^2dM_j^2}=\Sigma_q e_q^2\alpha_{em}\alpha_s^2
\pi^2[\xi G(\xi)]^2C\,\delta(M^2-M_j^2)\,\frac{16M^2Q^4\sqrt{1-\kappa}}
{3(M^2+Q^2)^6\kappa}\,.\label{mqq}
\ee
Here the $\delta$-function setting $M^2=M_j^2$ is only precise up to 
hadronization effects, which are expected to be of the order of the hadronic 
scale. The dependence on the transverse momentum cutoff enters via the 
variable $\kappa=4p'^2_{\perp,\mbox{\footnotesize cut}}/M_j^2$.

\gnufig{sp}{Distributions in $M^2$ and $M_j^2$ of diffractive events 
originating from \qq\ and \qqg\ final states for $Q^2=$50 GeV$^2$,   
$p'^2_{\perp,\mbox{\footnotesize cut}}=$5 GeV$^2$ and $C_g = 1$.}{
\setlength{\unitlength}{0.1bp}
\begin{picture}(3600,2160)(0,0)
\includegraphics{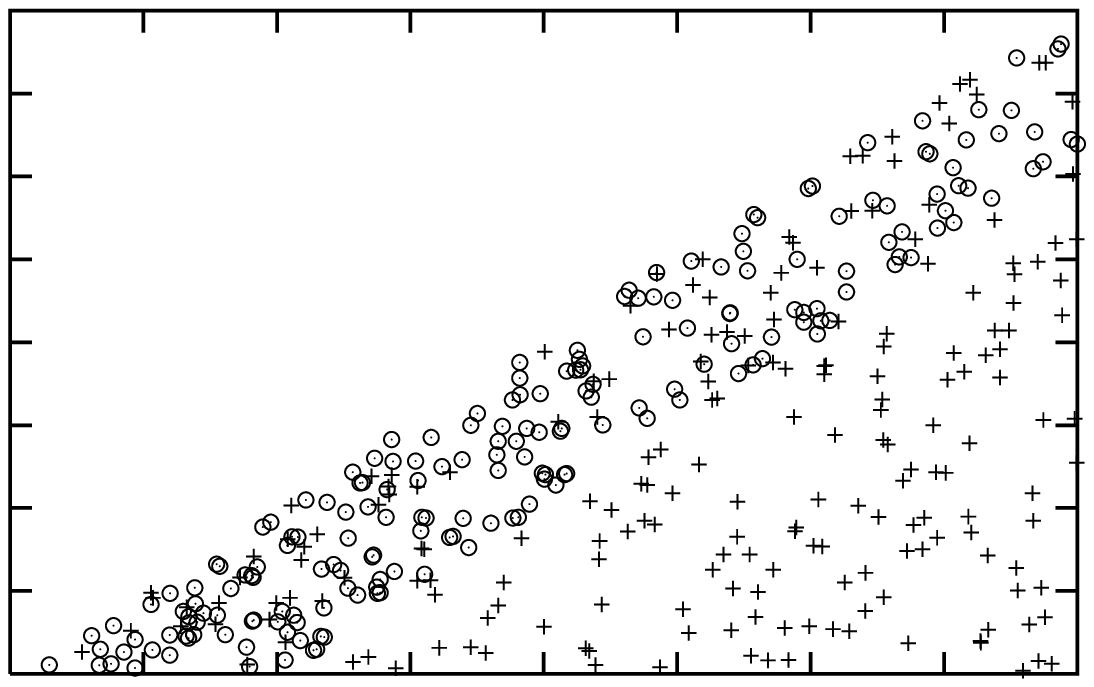}
\put(1132,1702){\makebox(0,0)[l]{$+ ~- ~q \bar{q} g$}}
\put(1132,1821){\makebox(0,0)[l]{$\odot ~- ~q \bar{q}  $}}
\put(1631,-136){\makebox(0,0)[l]{$M^2 ~(\mbox{GeV}^2)$}}
\put(-481,1105){\makebox(0,0)[l]{$M^2_{j} ~(\mbox{GeV}^2)$ }}
\put(3437,50){\makebox(0,0){100}}
\put(3053,50){\makebox(0,0){90}}
\put(2669,50){\makebox(0,0){80}}
\put(2284,50){\makebox(0,0){70}}
\put(1900,50){\makebox(0,0){60}}
\put(1516,50){\makebox(0,0){50}}
\put(1132,50){\makebox(0,0){40}}
\put(747,50){\makebox(0,0){30}}
\put(363,50){\makebox(0,0){20}}
\put(313,2060){\makebox(0,0)[r]{100}}
\put(313,1821){\makebox(0,0)[r]{90}}
\put(313,1583){\makebox(0,0)[r]{80}}
\put(313,1344){\makebox(0,0)[r]{70}}
\put(313,1105){\makebox(0,0)[r]{60}}
\put(313,866){\makebox(0,0)[r]{50}}
\put(313,628){\makebox(0,0)[r]{40}}
\put(313,389){\makebox(0,0)[r]{30}}
\put(313,150){\makebox(0,0)[r]{20}}
\end{picture}
}

In contrast, the mass distribution for diffractive processes with 
three-particle final states is not peaked at $M_j^2=M^2$. Concentrating, as 
before, on the transverse photon polarization and on the contribution from 
the diffractive gluon density the following formula can be derived from 
Eq.~(\ref{bgfu})
\be
\frac{d\sigma_T}{dM^2dM_j^2}=2\pi\Sigma_q e_q^2\alpha_{em}
\alpha_s\,y^2\frac{dg(y,\xi)}{d\xi}\,\,\frac{Q^4+M_j^4}{(Q^2+M_j^2)^5}\left[
2\,\mbox{Arctanh}\sqrt{1-\kappa}-\sqrt{1-\kappa}\,\right]\,.\label{mqqg}
\ee
Simple algebra shows that $y^2(dg(y,\xi)/d\xi)$, formally a function of $y$ 
and $\xi$, depends in fact only on the single variable $u=\xi/y-1$,
\be
y^2\frac{dg(y,\xi)}{d\xi}=\frac{1}{8u}\int \frac{d^2k_\perp'}{(2\pi)^4}
(k_\perp'^2)^2\int_{x_\perp}\left|\int\frac{d^2k_\perp}{(2\pi)^2}\left(
\delta^{ij}+\frac{2k_\perp^ik_\perp^j}{k_\perp'^2}u\right)\frac{\mbox{tr}
\tilde{W}^{\cal A}_{x_\perp}(k_\perp'\!-\!k_\perp)}{k_\perp'^2+uk_\perp^2}
\right|^2\,.
\ee
As has been argued in \cite{bhm2}, a simple parameterization of this 
function, consistent with the concept of a smooth localized colour field, is 
given by 
\be
y^2\frac{dg(y,\xi)}{d\xi} \propto \frac{1}{ C_g + u }\, ,
\ee
where $C_g$ is a constant of ${\cal O}(1)$. It has been checked that the 
precise value of the constant $C_g$ does not affect the qualitative features 
of the mass distribution (compare the analysis of \cite{bhm2}). 

To illustrate the experimental implications of Eqs.~(\ref{mqq}) and 
(\ref{mqqg}) we have shown in Fig.~\ref{fig:sp} the positions in $M^2$ and 
$M_j^2$ of two sets of 200 events, scattered randomly according to the  
distributions given above. The $\delta$-function of the \qq\ case has been 
replaced by a uniform distribution of $M_j^2$ in a band $M^2>M_j^2>M^2-20$ 
GeV$^2$. This allows for hadronization effects and, more importantly, for 
a large experimental uncertainty of the mass of the jet system. The
scatter plot clearly exhibits the distinctive features of the two
underlying partonic processes, even for this limited number of events.
With sufficient  statistics a determination of the relative weight of 
soft colour-singlet exchange, relevant in the \qqg\ case, and hard 
colour-singlet exchange, relevant in the \qq\ case, should be feasible.\\

\noindent
{\large\bf Conclusions}\\ \nopagebreak

The $p_\perp$-spectrum and the diffractive mass distribution for two 
different diffractive final states, \qq\ and \qqg, have been evaluated in 
the semiclassical approach. 

At high-$p_{\perp}$, a \qq\ final state is 
produced by hard colour-singlet exchange. The corresponding 
$p_{\perp}$-spectrum is identical with the result of two-gluon exchange in 
leading order. In contrast, high-$p_\perp$ jets in \qqg\ final states are 
predominantly produced via boson-gluon fusion. The colour neutralization 
mechanism is soft and the cross section is proportional to a diffractive 
gluon density. This non-perturbative quantity describes the probability of 
extracting a colour-singlet pair of low transverse momentum gluons from the 
proton.

The \qq\ and \qqg\ final states also lead to  qualitatively different
diffractive mass distributions. Hence, like diffractive charm production,
the study of diffractive high-$p_{\perp}$ jets should provide evidence
for the relative importance of soft and hard contributions in diffractive
deep-inelastic scattering.

\end{document}